\documentclass{iaus}
\usepackage{graphicx}

\title[feedback between SF and AGN] 
{Mutual feedback between star formation and nuclear activity}

\author[Gian Luigi Granato]   
{Gian Luigi Granato$^{1,2}$}%

\affiliation{$^1$INAF-OApd, Vicolo Osservatorio 5, Padova, Italy\\[\affilskip]
$^2$SISSA, Via Beirut 4, Trieste, Italy \break email:
gianluigi.granato@oapd.inaf.it}

\pubyear{2006}
\volume{235}  
\pagerange{119--126}
\date{?? and in revised form ??}
\jname{Galaxy Evolution across the Hubble
Time} \editors{F.Combes \& J. Palous, eds.}
\begin{document}

\maketitle

\begin{abstract}
In this invited contribution I review the justifications for the
attempts, currently very popular, to include in semi-analytic
models of galaxy formation prescriptions to describe the mutual
link between the star formation and nuclear activity in galaxies,
which has been for surprisingly long time neglected.
\end{abstract}

\firstsection 

\section{Introduction}
Since a few years ago, studies of galaxy formation have been
affected by uncertainties both in the cosmology and in the most
relevant physical processes. But now we are in the so called
"precision cosmology era", which implies that  the background
model is relatively well defined, and we can compute with
reasonable confidence and precision the evolution of the
dynamically dominant dark matter (DM) component, ruled essentially
only by gravity. By converse the main physical processes driving
galaxy formation, in particular the evolution of baryonic
(ordinary) matter, which allow us to see galaxies, are extremely
complex and still hotly debated.

Indeed, to compare scenarios and ideas of galaxy formation with
observations there are two strong sources of difficulties and
uncertainties.

\begin{itemize}

\item The first is to predict the evolution of ordinary matter,
which is highly non linear, and above all is driven mostly by
processes occurring well below the resolution of any feasible
simulation. These precesses are usually referred to as sub-grid
physics,  include star formation (SF), accretion onto
super-massive black holes (SMBH), merging of BH, and are also
poorly understood from a physical point of view.

\item Another point that has been for long time under-appreciated
is the modelling of  the interaction of photons produced by stars
and accretion processes with the dusty ISM, likely more and more
important at higher and higher z, to the point that it is likely
that a major fraction of SF activity at high z is completely
hidden to optical searches.

\end{itemize}

In this contribution I will discuss a few aspects of the first
point, though our group has devoted since long a lot of effort
also to cope the second problem, developing what is still the most
advanced and flexible tool to predict the spectral energy
distribution of galaxies in the presence of dust (GRASIL, Silva et
al.\ 1998; Vega et al.\ 2005).

\section{The myth of first principles models}

Due to the formidable task of computing the evolution of baryons
within galaxies in cosmological context, it is fair to say that
first principles or ab-initio models do not exist, despite some
optimistic claims. Indeed in any computation sub-grid physics is
treated (if ever) through semi-analytic-like recipes, that is
formulae containing many free parameters, which try to describe
sub-grid physics by means of relationships between integrated
quantities. Unfortunately, the model outcomes are heavily
affected, if not driven, by these formulae.

An illuminating example, related to the topic of this talk, are
the simulations of merging between disk galaxies carried by Di
Matteo et al.\ (2005). They included a rough sub-grid treatment of
the induced accretion onto a central SMBH and of the feed-back
that its activity may have on the evolution of the system.
Moreover, these are not simulations of cosmological volumes, thus
in principle sub-grid phenomena should be less a problem. However,
the striking result is that two comparison simulations, carried
respectively with and without the inclusion of these effects, lead
to completely different results in terms of star formation
histories and final fate of the initial gas. Without AGN the gas
is almost completely turned into stars, while with AGN feed-back
only about 50\% goes into stars and the rest is expelled from the
halo. Thus an ingredient which has to be treated with crude and
uncertain approximations dominates the predicted evolution. This
casts some perplexity on the rationale to attempt fully numerical
simulations of galaxy formation in cosmological volumes, and is
the basic justification of the wealth of semi-analytic models on
the market, in which all the processes involving baryons are
treated by means of simplified "recipes" or "prescriprions".

\section{Standard SAMs, their successes and their failures}

Indeed, the most extensive comparisons between different scenarios
and the rich data sets now available are done by means of fully
semi-analytic models (SAM) for baryons, possibly in the form of
post-processing of gravity-only simulations for DM. By definition,
these models are based on many a-priori assumptions on the general
development of galaxies. More proper naming could be toy models.

Almost all SAMs adopt a-priori what we can call  a disk galaxy
merger driven sequence of processes leading to galaxy populations
at any redshift and in particular locally. This is based mainly to
two assumptions: (i) that the first outcome of gas cooling in DMH
are disks of  gas supported by rotation (e.g.\ Rees \& Ostriker
1977, White \& Rees 1978), and in which there is only a mild SF
activity, while (ii) the only cause of violent episodes of SF at
any redshift are mergers between these disks, which are also the
main path to form spheroidal galaxies (e.g.\ White \& Frank 1991,
Cole et al.\ 2000).

These assumptions may be not completely correct. Indeed, despite
the high number of adjustable parameters involved, it has become
increasingly clear over the years that calculations based on this
general scheme, besides remarkable successes in reproducing
several properties of galaxy population, show severe mismatches
with some very basic observations. Namely: (i) without ad-hoc and
physically unjustified assumptions, it is impossible to reproduce
the sharp break at the bright end of the Local Luminosity
Function, in particular now that the low baryon fraction used in
old models has been ruled out by observations: models predict too
many (and too young-blue) bright galaxies (e.g. Benson et al
2003); (ii) by converse, beyond $z \sim 1-1.5$, most models tend
to predict too few bright galaxies, compared with the wealth of
determinations done in the past few years; (iii) these are aspects
of {\it cosmic downsizing}: star formation and accretion onto SMBH
decline more with cosmic time for larger system than for smaller
ones, a fact which is at odd with naive expectations of
hierarchical growth; (iv) the observed absence of cooling flows in
the centers of real rich clusters, where cooling times are much
shorter than the Hubble time; (v) the properties of local E
galaxies, such as the colors or the alpha enhancement as a
function of mass are not well reproduced. Most people think that
these difficulties may well be all facets of the same problem,
suggesting that some key ingredient is missing or/and the entire
scheme needs a substantial revision.

\section{Possible solution from joint evolution of QSO and Spheroids}

It is at present very popular the idea that at least part of the
solution could come from an ingredient that only in the very last
few years started to be taken into account in some models, i.e.\
the mutual influence or feed-back (FB) of star formation in
galaxies and the development of SMBH-QSO at their centers. This
influence is hinted by several empirical and also theoretical
facts, for instance the well established local correlation between
the mass of the central SMBH and several properties of the hosting
spheroid (the luminosity, the mass, the velocity dispersion), the
similarity of the cosmic development of SFR(z) and the luminosity
of QSO per unit volume, or the fact that simulations of merging
between galaxies drive flows of gas toward the central regions,
creating an environment at least favorable to promote SMBH
accretion.

Having now a relatively good determination of the almost constant
ratio of stellar mass in spheroids and the mass of the hosted SMBH
$\sim 1000$, it is interesting to compare the binding energy of
spheroids with the energy released by the SMBH to accrete this
mass. For a typical $L^*$ galaxy, the latter turns out to be more
than two orders of magnitude greater than the former, which means
that a quite small coupling between the energy released by the AGN
and the ISM is sufficient to have a significant effect. It is
interesting to compare also the binding energy with the energy
released by SNae during the assembly of the same spheroid, which
is only about a factor 10 greater than the former.

Then the energy available is large enough, though it is not clear
if and how a (small) fraction of this energy can be transferred to
the ISM. Several possibilities have been studied, in particular:
radiation pressure, mostly on dust grains: a dusty medium is
orders of magnitude more effective in absorbing momentum than a
dust-free one; Compton or line radiative heating; kinetic outflows
from AGN, such as those observed in jet and BAL (likely generated
by radiation pressure on resonant lines relatively close to the
central engine). Another proposed possibility is that the AGN may
act as a kind of catalyst to enhance the effectiveness of SNae
feed-back (Monaco 2004).

Only in the very last few years these effects started to be
explicitly considered in SAMs. This has been done along two quite
distinct lines that should not be confused.

\begin{itemize}
\item Granato et al.\ 2004, Monaco \& Fontanot 2005; Menci et al.\
2006 considered the FB associated with the main phase of the BH
growth, related to the bright high-z QSOs, as a way to sterilize
massive high-z galaxies, which instead are little affected by SNae
FB, due to the depth of their potential well.

\item More recently, and in the context of more standard SAMs, it
has been considered (only) FB associated with lower redshift, low
accretion rate phase of AGN, in which almost all the accretion
energy is used  to halt cooling flows and avoid overproduction of
local bright galaxies (Bower et al.\ 2006, Croton et al.\ 2006)
\end{itemize}

In general in this second set of works little attempt, or none at
all, has been done in treating the build up of SMBH and the
physical nature of the feedback.


\section{The ABC scenario}

The first SAM in which a key role has been invoked for the
reciprocal feed-back between star formation and AGN activity is
the "Antihierarchical Baryonic Collapse" (ABC) proposed by Granato
et al.\ 2004 (see also Granato et al.\ 2001 for a more
phenomenological treatment). This model, which is fully embedded
in the $\Lambda$CDM hierarchical growth of DM halos and is focused
on the formation of spheroids, adopts prescriptions to describe
the baryonic physics which reverse the order with which spheroidal
galaxies and high-z QSOs complete their formation, as indicated by
the various evidences of downsizing. This is obtained by a
combination of two ingredients: (i) revised prescriptions for the
SF in massive high redshift galactic halos ($M_{vir} \gtrsim
10^{12} M_{\odot}$), which allows SFR as high as thousands of
solar masses per year, as implied by observations of sub-mm
galaxies, and (ii) the inclusion of a treatment of the growth by
accretion of a SMBH promoted by this huge SF activity (positive
feed-back between SF and accretion), which at some point becomes
so powerful to clean the ISM and quench any further SF and
accretion (negative QSO feed-back).

The ABC scenario predicts a well defined evolutionary sequence
leading to local ellipticals with dormant SMBH. The sequence
begins with a high redshift phase of huge, dust enshrouded SF
activity best detectable in the sub—mm spectral region, and
lasting about 0.5 Gyr. This phase is ended by the strong feed-back
generated by the QSO phase, due to the growth of a SMBH, which is
promoted by the huge star formation activity during the previous
SMG phase. This feedback sterilizes the system, which then evolve
almost passively, thus it predicts a sizeable population of
massive and dead galaxies already at high-z. The model leads (in
one shot) to predictions in general agreement with many
observations which are at least disturbing for traditional SAMs



For the description of the model and how it compares with
observations of these populations of objects the reader is
referred to Granato et al.\ (2001, 2004), Silva et al.\ (2005);
Granato et al.\ (2006) Lapi et al.\ (2006); Silva et al.\ in
preparation.

\begin{acknowledgments}
The ideas presented in this invited contribution result from a
long standing collaboration with several colleagues, in
particular, listed in reverse alphabetic order,  L.\ Silva, F.\
Shankar, A.\ Lapi, G.\ De Zotti, L.\ Danese and A.\ Bressan. Work
in part supported by EC under CONTRACT MRTN-CT-2004-503929.
\end{acknowledgments}

\begin{discussion}
\end{discussion}


\begin{thebibliography}{}

\bibitem[\protect\citeauthoryear{Benson et al.}{2003}]{2003ApJ...599...38B}
Benson A.~J., Bower R.~G., Frenk C.~S., Lacey C.~G., Baugh C.~M.,
Cole S., 2003, ApJ, 599, 38

\bibitem[\protect\citeauthoryear{Bower et al.}{2006}]{2006MNRAS.370..645B}
Bower R.~G., Benson A.~J., Malbon R., Helly J.~C., Frenk C.~S.,
Baugh C.~M., Cole S., Lacey C.~G., 2006, MNRAS, 370, 645

\bibitem[\protect\citeauthoryear{Cole et al.}{2000}]{2000MNRAS.319..168C}
Cole S., Lacey C.~G., Baugh C.~M., Frenk C.~S., 2000, MNRAS, 319,
168

\bibitem[\protect\citeauthoryear{Croton et al.}{2006}]{2006MNRAS.365...11C}
Croton D.~J., et al., 2006, MNRAS, 365, 11

\bibitem[\protect\citeauthoryear{Di Matteo, Springel, \&
Hernquist}{2005}]{2005Natur.433..604D} Di Matteo T., Springel V.,
Hernquist L., 2005, Natur, 433, 604

\bibitem[\protect\citeauthoryear{Granato et
al.}{2001}]{2001MNRAS.324..757G} Granato G.~L., Silva L., Monaco
P., Panuzzo P., Salucci P., De Zotti G., Danese L., 2001, MNRAS,
324, 757

\bibitem[\protect\citeauthoryear{Granato et
al.}{2004}]{2004ApJ...600..580G} Granato G.~L., De Zotti G., Silva
L., Bressan A., Danese L., 2004, ApJ, 600, 580, (GDS04)

\bibitem[\protect\citeauthoryear{Lapi et al.}{2006}]{2006ApJ...650...42L}
Lapi A., Shankar F., Mao J., Granato G.~L., Silva L., De Zotti G.,
Danese L., 2006, ApJ, 650, 42

\bibitem[\protect\citeauthoryear{Menci et al.}{2006}]{2006ApJ...647..753M}
Menci N., Fontana A., Giallongo E., Grazian A., Salimbeni S.,
2006, ApJ, 647, 753

\bibitem[\protect\citeauthoryear{Monaco \&
Fontanot}{2005}]{2005MNRAS.359..283M} Monaco P., Fontanot F.,
2005, MNRAS, 359, 283


\bibitem[\protect\citeauthoryear{Rees \&
Ostriker}{1977}]{1977MNRAS.179..541R} Rees M.~J., Ostriker J.~P.,
1977, MNRAS, 179, 541

\bibitem[\protect\citeauthoryear{Silva et al.}{1998}]{1998ApJ...509..103S}
Silva L., Granato G.~L., Bressan A., Danese L., 1998, ApJ, 509,
103

\bibitem[\protect\citeauthoryear{Silva et al.}{2005}]{2005MNRAS.357.1295S}
Silva L., De Zotti G., Granato G.~L., Maiolino R., Danese L.,
2005, MNRAS, 357, 1295

\bibitem[\protect\citeauthoryear{Vega et al.}{2005}]{2005MNRAS.364.1286V}
Vega O., Silva L., Panuzzo P., Bressan A., Granato G.~L., Chavez
M., 2005, MNRAS, 364, 1286

\bibitem[\protect\citeauthoryear{White \& Rees}{1978}]{1978MNRAS.183..341W}
White S.~D.~M., Rees M.~J., 1978, MNRAS, 183, 341

\bibitem[\protect\citeauthoryear{White \&
Frenk}{1991}]{1991ApJ...379...52W} White S.~D.~M., Frenk C.~S.,
1991, ApJ, 379, 52

\end{thebibliography}
\end{document}